\documentclass[10pt,aps,prl,twocolumn,showpacs,amsmath,amssymb,superscriptaddress,floatfix]{revtex4}

\usepackage{bm}
\usepackage{graphicx}

\DeclareMathOperator{\str}{str}
\DeclareMathOperator{\sdet}{sdet}
\DeclareMathOperator{\diag}{diag}
\newcommand{\T}{\mathcal{T}}
\newcommand{\G}{\mathcal{G}}
\renewcommand{\H}{\mathcal{H}}
\newcommand{\U}{\mathrm{U}}
\renewcommand{\O}{\mathrm{O}}
\newcommand{\OSp}{\mathrm{OSp}}
\newcommand{\Sp}{\mathrm{Sp}}
\newcommand{\SpO}{\mathrm{SpO}}
\newcommand{\SO}{\mathrm{SO}}
\newcommand{\GL}{\mathrm{GL}}
\newcommand{\Z}{\mathbb{Z}}

\makeatletter
\def\NAT@bibsetnum#1{
 \setlength{\topsep}{\z@}
 \NATx@bibsetnum{#1}
}
\renewenvironment{thebibliography}[1]{
 \NAT@thebibliography{#1}
 \@clubpenalty\clubpenalty
 \let\@TBN@opr\present@bibnote
 \@FMN@list
}{
 \edef\@currentlabel{\arabic{NAT@ctr}}
 \NAT@endthebibliography
 \global\let\auto@bib\@empty   
}
\newcommand*{\supplementarystart}{
  \close@column@grid
  \clearpage
  \onecolumngrid
  \setcounter{enumiv}{0} 
  \setcounter{equation}{0} 
  \setcounter{figure}{0} 
  \setcounter{table}{0} 
  \setcounter{page}{1}
  \c@secnumdepth=4
  \renewcommand{\theequation}{S\arabic{equation}} 
  \renewcommand{\bibnumfmt}[1]{[S##1]} 
  \renewcommand{\@onlinecite}{S\citealp} 
  \renewcommand{\cite}[1]{{[}\onlinecite{##1}{]}}
  \renewcommand{\thefigure}{S\arabic{figure}}
  \renewcommand{\thetable}{S\Roman{table}}
  \renewcommand{\thepage}{S\arabic{page}}
}
\makeatother

\begin{document}

\title{Dynamics of Anderson localization in disordered wires}
\author{E.\ Khalaf}
\affiliation{Max Planck Institute for Solid State Research, Heisenbergstr.\ 1, 70569 Stuttgart, Germany}
\author{P.\ M.\ Ostrovsky}
\affiliation{Max Planck Institute for Solid State Research, Heisenbergstr.\ 1, 70569 Stuttgart, Germany}
\affiliation{L.\ D.\ Landau Institute for Theoretical Physics RAS, 142432 Chernogolovka, Russia}

\begin{abstract}
We consider the dynamics of an electron in an infinite disordered metallic wire. We derive exact expressions for the probability of diffusive return to the starting point in a given time. The result is valid for wires with or without time-reversal symmetry and allows for the possibility of topologically protected conducting channels. In the absence of protected channels, Anderson localization leads to a nonzero limiting value of the return probability at long times, which is approached as a negative power of time with an exponent depending on the symmetry class. When topologically protected channels are present (in a wire of either unitary or symplectic symmetry), the probability of return decays to zero at long time as a power law whose exponent depends on the number of protected channels. Technically, we describe the electron dynamics by the one-dimensional supersymmetric non-linear sigma model. We derive an exact identity that relates any local dynamical correlation function in a disordered wire of unitary, orthogonal, or symplectic symmetry to a certain expectation value in the random matrix ensemble of class AIII, CI, or DIII, respectively. The established exact mapping from one- to zero-dimensional sigma model is very general and can be used to compute any local observable in a disordered wire.
\end{abstract}


\maketitle

\emph{Introduction.---}
Quantum interference leads to localization of electrons in the presence of disorder. In one- (1D) and two-dimensional (2D) systems, even weak random potential localizes all eigenstates, while in three dimensions (3D) localization occurs when disorder is stronger than a certain threshold level \cite{Mott61, Thouless74, Abrahams79}. In the past few years, the phenomenon of Anderson localization has witnessed a revival of activity due to discoveries made in several fields. On the experiment side, Anderson localization has been observed in a multitude of systems including cold atoms \cite{Billy08, Roati08, Aspect09}, light waves \cite{Wiersma97}, ultrasound \cite{Faez09}, as well as optically driven atomic systems \cite{Lemarie10}. On the theory side, dynamical phenomena such as thermalization and relaxation after a quantum quench in disordered systems have been the subject of growing interest \cite{Ziraldo12, Ziraldo13, Rahmani15, Canovi11, Bardarson12}. This has been inspired, in part, by the discovery of many-body localization \cite{Altshuler97, Gornyi05, Basko06, Pal10, Nandkishore15}, which is an interacting analog of Anderson localization, and more recently by the proposal to diagnose quantum chaotic behavior by means of out-of-time-order correlations \cite{Aleiner16, Chen16, Bagrets17, Fan17, Swingle17}. Furthermore, the discovery \cite{Kane05a, Kane05b, Bernevig06, Koenig07, Koenig08, Hasan10, Moore09} and complete classification \cite{Altland97, Schnyder08, Schnyder09, Kitaev09, Ryu10} of topological insulators has opened the door to a new arena where the interplay between disorder and topology leads to unusual localization-related effects. These include ultra-slow (Sinai) diffusion at the critical phase between two topological insulator phases \cite{Bagrets16}, as well as enhanced localization effects in systems where topologically protected and unprotected channels coexist \cite{KhalafSC, KhalafA}.

Despite more than half a century since Anderson's original paper \cite{Anderson58}, there exists very few exact results \cite{Skvortsov07} about electron dynamics in the Anderson-localized phase beyond the strictly 1D (single channel) case \cite{Gorkov83}. In particular, the absence of exact results for dynamical correlations in disordered wires (quasi-1D multichannel system) is rather surprising in light of the remarkable success of the field theoretic approach to the problem in terms of the supersymmetric non-linear sigma model (NLSM). The NLSM method has proven to be very efficient in describing static response \cite{Efetov83, Efetov99, Mirlin00} and have been successfully employed to obtain the conductance, its mesoscopic fluctuations \cite{Zirnbauer92, Mirlin94, Efetov99}, as well as the full distribution function of transmission eigenvalues \cite{Rejaei96, Lamacraft04, Altland05, KhalafA} in disordered wires. In addition to being an effective model for localization problems in general, NLSM is a generic field theory arising in a number of other problems such as random banded matrices \cite{Fyodorov91, Mirlin00} and the dynamics of the quantum kicked rotor \cite{Casati79, Fishman82, Grempel84,Izrailev90, Altland96}.

In this Letter, we provide an exact analytic expression for an arbitrary local dynamical correlation (LDC) function of a disordered metallic wire in one of three Wigner-Dyson symmetry classes. This is done by showing that, rather surprisingly, any LDC of the supersymmetric 1D NLSM in the unitary, orthogonal, or symplectic class is given \emph{exactly} by a corresponding correlation function of a zero-dimensional NLSM in one of the classes AIII, CI, and DIII, respectively. The latter can always be evaluated explicitly as a finite dimensional integral.

The result is quite general and can be used to compute any LDC such as correlations of the local density of states at different energies, out-of-time-order correlations of operators at nearby points, and diffusion probability of return. We will focus on the latter quantity since it is the simplest to compute and most intuitive to understand. It is also readily observable in time-resolved measurements of the electron density profile, which is possible e.g.\ in cold atom experiments \cite{Billy08, Roati08, Aspect09}. We will consider the possibility of having topologically protected channels coexisting with regular channels in the quasi-1D wire. This could be realized in the vicinity of a doped Weyl point in magnetic field \cite{Altland16, KhalafA} or at the edge of a 2D topological or Chern insulator \cite{KhalafSC}.

\emph{Formalism.---}
We consider a model of an infinite quasi-1D metallic wire with $N \gg 1$ conducting channels with or without time reversal symmetry (TRS) $\T$. The system belongs to one of three Wigner-Dyson symmetry classes: unitary (no TRS), orthogonal ($\T^2 = 1$), or symplectic ($\T^2 = -1$). In the absence of TRS, the numbers of left- and right-moving channels generally differ by an integer $m$ that represents a topological invariant and corresponds to the number of chiral topologically protected channels. The presence of TRS enforces the number of left- and right-moving channels to be the same. In this case, it is possible to have a single helical topologically protected channel if $\T^2 = -1$ (symplectic class) and the total number of channels $N$ is odd.

Any LDC of a disordered system can be expressed as the disorder-averaged product of Green's functions. Dynamical correlations involve Green's functions at two different energies, whereas local correlations involve Green functions between spatially close points within the localization length $\xi = N l$, where $l$ is the mean-free path. The main quantity we will consider in this work is the return probability $W(t)$, which is the probability that a diffusing electron returns to the starting point after time $t$. It can be expressed in terms of the disorder average of two Green's functions as
\begin{equation}
\label{Wt1}
 W(t)
  = \int \frac{d\omega \, e^{-i \omega t}}{4 \pi^2 \nu} \bigl< G^R_{\epsilon + \omega}(x, x') G^A_{\epsilon}(x', x) \bigr> \Bigr|_{x' \to x},
 \end{equation}
with $\nu$ being the density of states. The limit $x' \to x$ implies that $l \ll |x' - x| \ll \xi$; the first inequality excludes any nonuniversal ballistic effects.
 
Disorder averaging of a product of Green's functions can be performed following the standard procedure \cite{Efetov83, Efetov99, Mirlin00} that starts by writing this product as a Gaussian integral over supervector field $\psi$. Averaging over disorder leads to a quartic term in $\psi$ that is decoupled with the help of a supermatrix field $Q$. The effective field theory in terms of $Q$ is obtained by means of a saddle point approximation followed by a gradient expansion. 
 
The resulting action at an imaginary frequency $\omega = i \Omega$ has the form of a non-linear sigma model \cite{Efetov83, Efetov99, Mirlin00, KhalafSC, KhalafA}
\begin{equation}
\begin{aligned}
 &S = -\frac{\pi\nu}{4\gamma} \int dx \str \left[ D (\partial_x Q)^2 - 2 \Omega \Lambda Q \right] + S_\text{top}, \\
 &S_\text{top} = \frac{m}{2} \int dx \str \bigl( T^{-1} \Lambda \partial_x T \bigr), \qquad Q = T^{-1} \Lambda T.
\end{aligned}
\label{SQ}
\end{equation}
Here $D$ is the diffusion constant and $\gamma$ is given in Table \ref{table}. The topological term $S_\text{top}$ involves an integer number $m$ denoting the difference between the number of left- and right-moving channels in a unitary wire, or the total number of channels in a symplectic wire. The matrices $T$ and $Q$ operate in the direct product of retarded-advanced (RA), Bose-Fermi (BF), and (if TRS is present) time-reversal (TR) spaces in addition to the space of $n$ replicas. The latter is required to compute an average of $2n$ Green's functions \cite{footnote1}. The matrix $\Lambda$ is $\diag\{1,-1\}_\mathrm{RA}$.
 
\begin{table}
\centerline{\begin{tabular}{cccccc}
 \hline\hline
 Class & $\gamma$ & $\G(n)$ & noncompact & compact & Topology \\
 \hline
 Unitary & $1$ & AIII & $\mathrm{GL}(n, \mathbb{C})/\mathrm{U}(n)$ & $\mathrm{U}(n)$ & $\mathbb{Z}$ \\
 Orthogonal & $2$ & CI & $\mathrm{SO}(2n,\mathbb{C})/\mathrm{SO}(2n)$ & $\mathrm{Sp}(2n)$ &$0$ \\
 Symplectic & $2$ & DIII & $\mathrm{Sp}(2n,\mathbb{C})/\mathrm{Sp}(2n)$ & $\mathrm{O}(2n)$ & $\mathbb{Z}_2$ \\
 \hline\hline
\end{tabular}}
\caption{Sigma-model manifolds for Wigner-Dyson classes $Q \in \G(2n)/\G(n) \times \G(n)$. The parameter $\gamma$ accounts for the size of the matrix and normalizes the supertraces. The effective 0D sigma model defined on the group manifold $\G(2n)$ is used in the integral representation (\ref{FQ3}).}
\label{table}
\end{table}

The matrix $T$ is an element of a Lie supergroup $\G(2n)$ given in Table \ref{table} for the three classes \cite{footnote2}. The matrix $Q$, parametrized as $T^{-1} \Lambda T$, is invariant under left multiplication $T \mapsto KT$ by any matrix $K$ that commutes with $\Lambda$. As a result, $Q$ belongs to the coset space $\G(2n)/\G(n) \times \G(n)$ \cite{footnote3}. We restrict $T$ and $K$ to have unit superdeterminant $\sdet T = \sdet K = 1$, which is necessary for the proper definition of $S_\text{top}$ in Eq.\ (\ref{SQ}) \cite{KhalafSC}.

The topological term $S_\text{top}$ is not invariant under gauge transformations $T \mapsto KT$ but rather changes by an integral of a total derivative, much like the action of a charged particle in an external magnetic field \cite{KhalafSC}. In the three symmetry classes, the value of $S_\text{top}$ is either identically zero (orthogonal), $0$ and $i \pi$ (symplectic), or an arbitrary imaginary number (unitary). Hence the value of $m$ is immaterial in an orthogonal wire. In symplectic wires, only the parity of $m$ is relevant distinguishing the cases of even and odd number of channels. In the unitary class, $m$ corresponds to the imbalance between left- and right-moving channels.

\emph{Evolution operator and correlation functions.---}
Any LDC is expressed in the sigma-model language as the expectation value of a function of $Q$ at a single point
\begin{equation}
\label{FQ1}
 \langle F(Q) \rangle
  = \int \mathcal{D}Q\, F[Q(x=0)] e^{-S[Q]},
  \end{equation}
with the action $S[Q]$ given by Eq.\ (\ref{SQ}). In particular, the return probability $W(t)$, defined in Eq.\ (\ref{Wt1}), can be written as
\begin{equation}
  \label{Wt2}
  W(t) = -\nu \!\int\! \frac{d\omega\, e^{-i \omega t}}{16\gamma^2} \mathop{\mathrm{str}} \bigl< k P_+ Q k P_- Q\bigr>, \quad P_\pm = \frac{1 \pm \Lambda}{2}.
\end{equation}
Here $k = \diag\{1,-1\}_\mathrm{BF}$ is the grading matrix.

Calculation of the expectation value (\ref{FQ1}) is facilitated by defining the evolution operator, that is a path integral on the half-infinite wire
\begin{equation}
 \label{psi}
 \psi_m(T) = \int_{x=0,Q(0)=T^{-1} \Lambda T}^{x=\infty, Q(\infty) = \Lambda} \mathcal{D}Q  \, e^{-S[Q]}.
\end{equation}
We write the evolution operator as a function of $T$ rather than $Q$ to emphasize its gauge dependence. Under a gauge transformations $T \mapsto K T$, it transforms as
\begin{equation}
\label{gauge}
\psi_m(K T) = (\sdet K_R)^m \psi_m(T) = (\sdet K_A)^{-m} \psi_m(T)
\end{equation}
in full analogy to a wave function in magnetic field.

In Eq.\ (\ref{gauge}), $K_{R/A}$ are the two (retarded and advanced) blocks of the matrix $K$, each from the supergroup $\G(n)$. The restriction $\sdet K = 1$ ensures the equivalence of the two expressions in Eq.\ (\ref{gauge}). The product $\psi_m(T) \psi_{-m}(T)$ is gauge invariant and hence depends on $Q$ only. This allows us to write the expectation value $\langle F(Q) \rangle$ as an ordinary rather than path integral:
\begin{equation}
\label{FQ2}
\langle F(Q) \rangle = \int dQ \, \psi_{-m}(T) F(Q) \psi_m(T).
\end{equation}

The function $\psi_m(T)$ can be identified with the zero mode of the transfer matrix Hamiltonian corresponding to the action (\ref{SQ}) with the coordinate $x$ playing the role of a fictitious imaginary time. Under evolution in $x$, all nonzero modes exponentially decay hence only the zero mode survives in a half-infinite wire. The transfer matrix Hamiltonian contains a kinetic term, represented by the Laplace-Beltrami operator on the sigma model manifold, and a potential term $\str{\Lambda Q}$ \cite{SuppMat}.

The main result of this Letter is an explicit integral representation of $\psi_m(T)$ that we construct as
\begin{equation}
\label{psi0}
\psi_m(T) = \int
dK \left(\sdet K_R \right)^m \phi(K T).
\end{equation}
This integral runs over $K \in \G(n)\times\G(n)$ constrained by $\sdet K = 1$. For any function $\phi(T)$, the above integral represents an average over the gauge group $K$ with the weight $(\sdet K_R)^m$ that ensures the correct transformation properties (\ref{gauge}). Choosing the function $\phi$ to be
\begin{equation}
\label{phi0}
\phi(T) = \exp\left[-\frac{\kappa}{2\gamma} \str P_\pm (T + T^{-1})\right],
\quad
\kappa = 4\pi\nu \sqrt{D\Omega},
\end{equation}
we observe that the integral (\ref{psi0}) is annihilated by the transfer matrix Hamiltonian, which is shown explicitly in the supplemental material \cite{SuppMat}, and hence indeed provides an explicit expression for the zero mode.

Several comments are in place here about the expression for the evolution operator [Eqs.\ (\ref{psi0}) and (\ref{phi0})]. First, the integral (\ref{psi0}) can be equivalently written with the factor $(\sdet K_A)^{-m}$, while the function $\phi(T)$ contains any of the two projection operators $P_\pm$ defined in Eq.\ (\ref{Wt2}). It turns out that the result of integration is independent of the choice of $P_\pm$. In both cases, integration over $K$ in Eq.\ (\ref{psi0}) reduces to integration over $K_R$ or $K_A$ within the group $\G(n)$, since the integrand depends only on one of the two blocks of $K$. Second, the very existence of the zero mode relies crucially on the supersymmetry. Both compact and non-compact replica sigma models do not possess a zero mode and the function defined in Eqs.\ (\ref{psi0})--(\ref{phi0}) \emph{does not} vanish under the action of the transfer matrix Hamiltonian. However, the result of such an action \emph{does} vanish in the replica limit $n \to 0$. This means that the simple integral representation for the evolution operator is an exclusive feature of symmetric superspaces not shared by their compact or non-compact non-supersymmetric counterparts. Third, the expression (\ref{psi0}) already captures correct topological properties of the three classes. The determinant factor is always $1$ in the orthogonal class and thus drops out for any $m$, while it equals $\pm 1$ in the symplectic class making it sensitive only to the parity of $m$. In the unitary class, the determinant represents a phase factor and hence distinguishes all integer values of $m$.

An arbitrary LDC can now be expressed using Eqs.\ (\ref{FQ2}), (\ref{psi0}), and (\ref{phi0}). The integral for $\langle F(Q) \rangle$ contains the functions $\psi_m$ and $\psi_{-m}$. We choose the form with $K_R$ integral in (\ref{psi0}) for one of them and with the $K_A$ integral for the other. This amounts to using two different projectors $P_\pm$ for the two functions. The integrals over $Q$, $K_R$, and $K_A$ can be combined into a single integral over $T \in \G(2n)$ leading to the remarkably simple expression
\begin{multline}
\label{FQ3}
\langle F(Q) \rangle =  \! \! \int_{\G(2n)} \! \! dT\, (\mathop{\mathrm{sdet}} T)^m F(T^{-1} \Lambda T) \\
   \times \exp\left[ -\frac{\kappa}{2\gamma} \mathop{\mathrm{str}} ( T + T^{-1} )\right],
\end{multline}
where the assumption $\sdet T=1$ has been dropped.

Integrals of the type (\ref{FQ3}) were previously studied in the context of Gaussian ensembles of random chiral matrices \cite{Ivanov02}. Equation (\ref{FQ3}) relates any local correlation function of a 1D sigma model at frequency $\Omega$ to the correlation function of a 0D sigma model at frequency $\kappa \sim \sqrt{\Omega}$ \emph{in a different symmetry class}. The unitary, orthogonal, and symplectic classes map to classes AIII, CI, and DIII, respectively, see Table \ref{table}.

\emph{Return probability.---}
We now demonstrate the power of Eq.\ (\ref{FQ3}) and compute the return probability Eq.\ (\ref{Wt2}). We employ the minimal $n = 1$ model and use a specific parametrization of $T \in \G(2)$ whose details are given in the supplemental material \cite{SuppMat}. The result takes the simplest form in terms of the inverse dimensionless time $z$:
\begin{equation}
W(t) = \frac{F(z)}{8\pi\nu D},
\qquad z = \frac{1}{\tau} = \frac{8\pi^2\nu^2 D}{t}.
\end{equation}
The function $F(z)$ is given by
\begin{subequations}
\label{Ft}
\begin{align}
 &\!\!F^\text{U}_m(z)
  = \frac{2 e^{-z}}{3} \Bigl[ (2z + m + 2) I_m(z) + z I_{m + 1}(z) \Bigr], \label{FU}\\
&\!\!F^\text{O}(z)
  = 1 + \frac{e^{-z}}{3} \Bigl[ (3z + 5) I_0(z) + (3z + 4) I_1(z) \Bigr], \label{FO} \\
   &\!\!F^\text{Sp}_\text{e/o}(z)
  = F^\text{O}(z) - 2 \pm \frac{e^{-z/2}}{3} (z + 2).\label{FS} 
\end{align}
\end{subequations}
Here $I_m(z)$ denotes the modified Bessel function. These simple expressions capture the complete cross-over between classical diffusion at short times $\tau \ll 1$ and localization at long times $\tau \gg 1$.

\begin{figure}
\center
\includegraphics[width=0.45\textwidth]{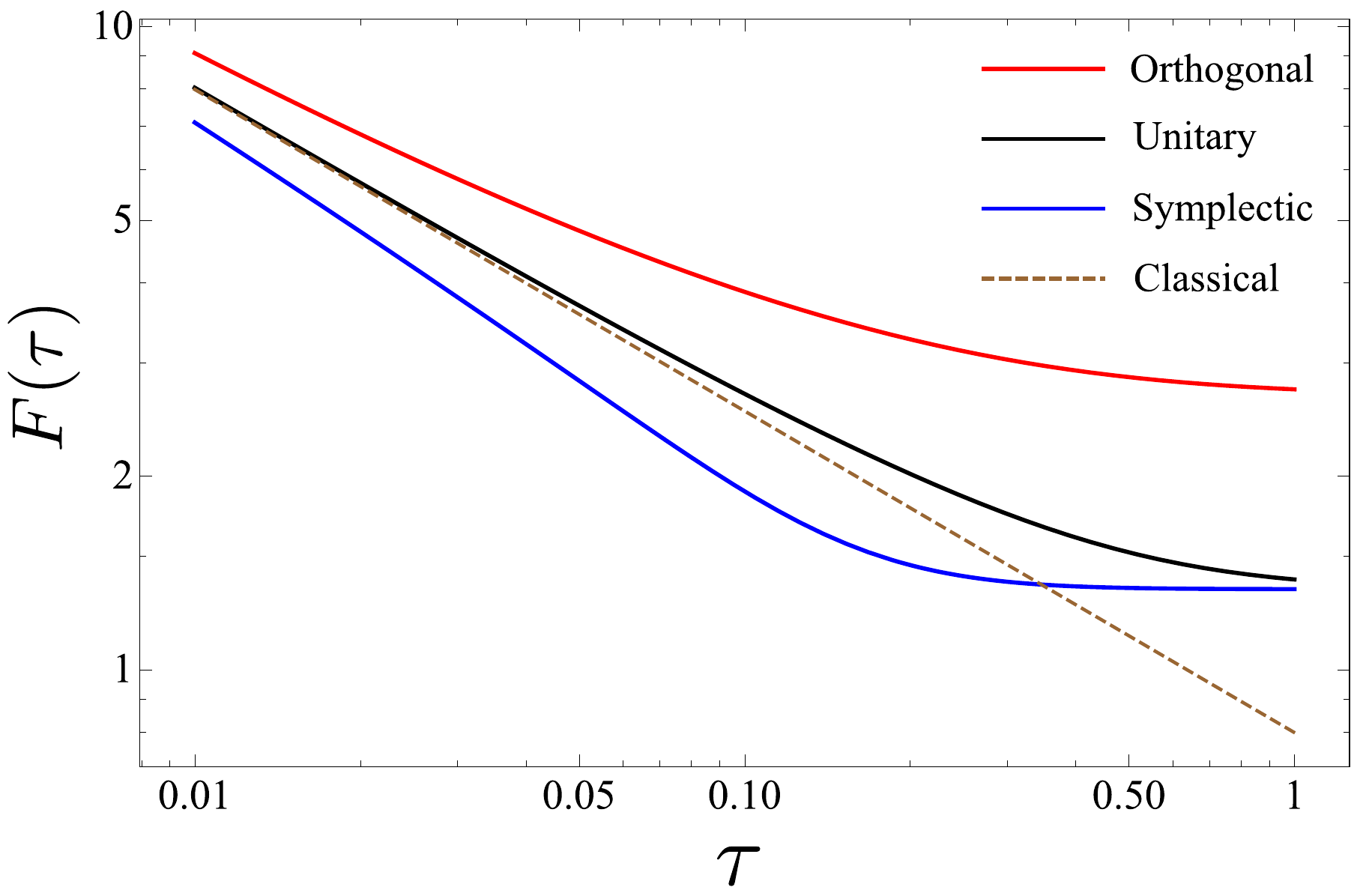}
\caption{Return probability $F(\tau)$ as a function of dimensionless time $\tau$ (log scale) for the unitary (black), orthogonal (red) and symplectic (blue) classes together with the result for classical diffusion (dashed).}
\label{Fdm}
\end{figure}

The return probability $F(\tau)$ in the absence of any topological channels is plotted in Fig.\ \ref{Fdm}. At short times $\tau \ll 1$, all the curves approach the result for classical diffusion $F = \sqrt{2/\pi \tau}$. The leading correction to the classical result is given by $\pm 1$ for the orthogonal/symplectic class indicating weak localization/antilocalization. In the unitary class, localization correction $(5/4)\sqrt{\tau/2\pi}$ appears only in the second order.

At long times $\tau \gg 1$, all curves approach a non-zero saturating value indicating localization. This value is $4/3$ for the unitary and symplectic classes and $8/3$ for the orthogonal class. This is consistent with the fact that the localization length in the latter case is twice shorter \cite{Evers08, Mirlin94}. The function $F(\tau)$ approaches its saturation value as a power law $\sim 1/\tau^3$, $1/\tau^2$, and $1/ \tau^5$ in the unitary, orthogonal, and symplectic classes, respectively.

\begin{figure}
\center
\includegraphics[width=0.45\textwidth]{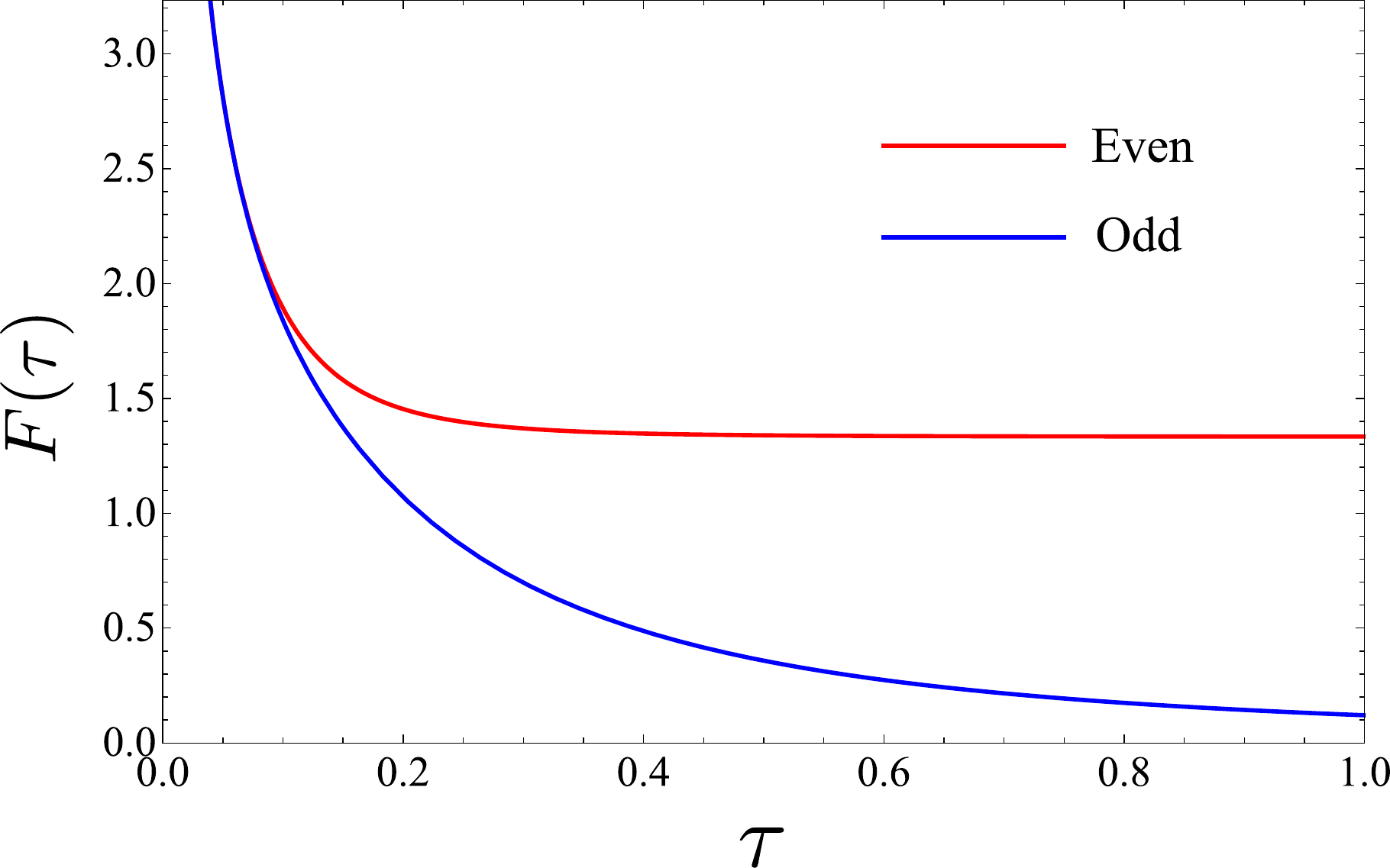}
\caption{Return probability $F(\tau)$ in a symplectic wire with an even (red) and odd (blue) total number of channels.}
\label{Fsm}
\end{figure}

Return probabilities in symplectic wires with even and odd number of channels are compared in Fig.\ \ref{Fsm}. Short-time asymptotics of $F(\tau)$ is insensitive to the parity to all orders. This shows that the effects of $\Z_2$ topology are invisible on the perturbative weak localization level \cite{KhalafSC}. At long times, the curve for odd number of channels decays to zero as $\sim 1/\tau^2$ indicating delocalization due to the presence of a single topologically protected channel.

\begin{figure}
\center
\includegraphics[width=0.45\textwidth]{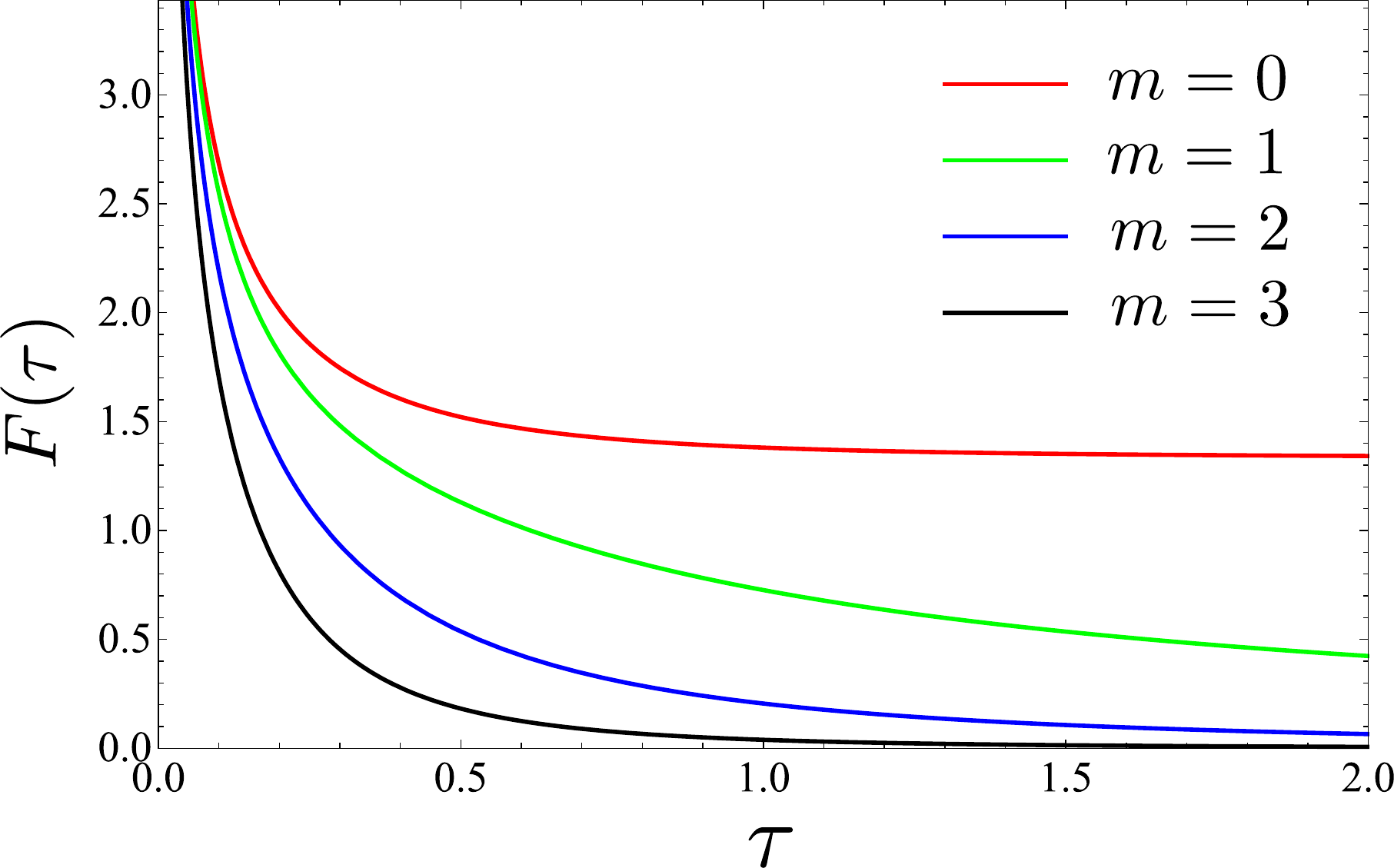}
\caption{Return probability $F(\tau)$ in a unitary wire for several different values of the channel imbalance $m$.}
\label{Fum}
\end{figure}

Return probability in a unitary wire is shown in Fig.\ \ref{Fum} for different values of the channel imbalance $m$. For $m \neq 0$, the curves decay to zero as $\sim 1/\tau^m$ indicating delocalization. The decay power increases with $m$ since the delocalization enhances with increasing the number of topologically protected channels. It is instructive to compare this result to the classical picture of diffusion accompanied with a unidirectional drift due to protected chiral channels \cite{KhalafSC, Altland16}. In the classical limit, the return probability is given by $F(\tau) = \sqrt{2/\pi \tau} e^{-m^2 \tau/2}$ and decays exponentially at long times. This corresponds to a Gaussian wave packet that spreads as $\sqrt{2 D t}$ and drifts with a constant velocity $m/2\pi\nu$. Localization corrections turn the exponential decay of return probability into a power law indicating that the drifting wave packet leaves a ``fat tail'' behind.

\emph{Discussion and conclusion.---}
The main result of this Letter is the identity (\ref{FQ3}) that relates an arbitrary local correlation function of the 1D NLSM at finite frequency to the correlation function of a 0D NLSM in a different symmetry class. The latter can be evaluated explicitly as a finite dimensional integral. The result applies to supersymmetric models with an arbitrary number of replicas, is valid for disordered metallic wires in the presence or absence of time-reversal symmetry, and allows for an arbitrary topological index $m$.  It remains to be seen whether the result can be generalized further to superconducting and chiral symmetry classes. The exact identity between correlation functions of the 1D and 0D NLSM raises an intriguing possibility that similar relations could also hold in higher dimensions.

The identity (\ref{FQ3}) was applied to study diffusion probability of return, which is the simplest local dynamical observable. We obtained exact analytic expressions (\ref{Ft}) that cover the complete crossover from the short-time semiclassical (weak localization) regime to the long-time strong localization regime. The return probability has a nonzero value at long times indicating complete localization in wires without topologically protected channels  (Fig.\ \ref{Fdm}). This saturation value is approached as a power law in time with an exponent that depends on the symmetry class. In the presence of protected channels, the return probability decays to zero as a power-law in time (Figs.\ \ref{Fsm} and \ref{Fum}) with an exponent that depends on the topological index $m$. This power-law decay arises due to quantum interference effects and is in sharp contrast with the exponential decay predicted by the classical model of diffusion and drift.

The general result (\ref{FQ3}) can be used to compute various physical observables in disordered systems \textit{exactly}. In addition to the diffusion probability of return considered here, these observables include out-of-time-order correlations \cite{Swingle16}, correlations of the local density of states at different energies \cite{Skvortsov07} (which can be probed in optical response experiments), zero-bias anomaly in disordered wires in the presence of short-range interactions \cite{AltshulerAronov}, strong Anderson localization peak in cold atom quantum quenches \cite{Karpiuk12, Micklitz14, Micklitz15}, as well as the proximity effect at the interface between a superconductor and a disordered wire \cite{Skvortsov13, Ivanov14}.

\emph{Acknowledgments.---}
We are grateful to D.\ Bagrets, I.\ Gornyi, E.\ K\"onig, A.\ Mirlin, I.\ Protopopov, and M.\ Skvortsov for valuable discussions. The work was supported by the Russian Science Foundation (Grant No.\ 14-42-00044).

\supplementarystart

\centerline{\bfseries\large ONLINE SUPPLEMENTAL MATERIAL}
\vspace{6pt}
\centerline{\bfseries\large Dynamics of Anderson localization in disordered wires}
\vspace{6pt}
\centerline{E.\ Khalaf and P.\,M.\ Ostrovsky}
\begin{quote}
In this Supplemental Material, we provide technical details relevant for the text of the Letter. First, we present the transfer matrix Hamiltonian corresponding to the sigma-model action (2) and show that the function defined by Eqs.\ (8) and (9) is annihilated by the action of this Hamiltonian. Second, we construct the zero mode explicitly in the minimal (one replica) model for the unitary, orthogonal, and symplectic classes using the integral representation (8)--(9). Finally, we calculate the return probability (12) from the general expression for local correlation functions (10).
\end{quote}

\section{Evolution operator}

In this section, we show that the evolution operator defined in (5) is indeed given by the integral representation (8) and (9). We will first present the transfer matrix Hamiltonian corresponding to the sigma-model action (2). We will then derive a set of identities that allows us to simplify expressions involving a sum over generators of a Lie (super)algebra in terms of (super)traces of operators. These identities will be crucial to evaluate the action of the transfer matrix Hamiltonian on the function defined by Eqs.\ (8) and (9) and to show that it indeed yields the zero mode.

\subsection{Transfer matrix Hamiltonian}

Our starting point is the evolution operator (5). In order to proceed further, we make the observation that the one-dimensional path integral with the sigma-model action is equivalent to the quantum mechanical evolution operator with the position $x$ playing the role of imaginary time. As a result, the evolution operator $\psi$ on the sigma model manifold at the point $x$ (here we measure the distance in units of localization length $\xi$) satisfies the Schr\"odinger equation
\begin{equation}
\label{Schr}
 \frac{\partial \psi(T,x)}{\partial x} = -\H \psi(T,x).
\end{equation}
Here, as in Eq.\ (5), the evolution operator $\psi$ is written as a function of $T$ rather than $Q$ to stress its gauge dependence, which follows from the gauge dependence of the action. 

The transfer matrix Hamiltonian (up to an unimportant constant factor) has the form
\begin{equation}
\label{H}
 \H = -\frac{\gamma}{2} \Delta_Q + \frac{\kappa^2}{16 \gamma} \str(\Lambda Q).
\end{equation}
Here $\Delta_Q$ is the Laplace-Beltrami operator on the sigma-model supermanifold $Q \in \G(2n)/\G(n) \times \G(n)$. It should be stressed that the Laplace-Beltrami operator acts on the coset space of the matrix $Q$ rather than the bigger manifold $\G(2n)$ for the matrix $T$. The action of $\Delta_Q$ on a function of $T$, which transforms according to Eq.\ (6), can be expressed by introducing local coordinates on the coset space. This can be achieved by choosing the set of generators $t_i$ of the Lie superalgebra of $\G(2n)$ that anticommute with $\Lambda$. The action of the Laplace-Beltrami operator is then given by
\begin{equation}
\label{Delta}
 \Delta_Q \psi(T) = \eta^{ij} \frac{\partial}{\partial x^i} \frac{\partial}{\partial x^j} \psi\left( e^{X/2} T \right)\Bigr|_{X = 0},
 \qquad X = x^l t_l,
 \qquad \eta_{ij} = \str(t_i t_j),
 \qquad \eta_{ij} \eta^{jk} = \delta_i^k.
\end{equation}
In these expressions, summation over repeated Latin indices (one lower and one upper) is implied and $\eta^{ij}$ is a matrix inverse of $\eta_{ij}$. It is easy to see that the action of the operator defined in (\ref{Delta}) preserves the transformation relation (6).

It follows from Eq.\ (\ref{Schr}) that the evolution operator between two points can be expanded in eigenfunctions of the transfer matrix Hamiltonian. Each term in this expansion decays with $x$ as $e^{-x \epsilon}$, where $\epsilon$ is the corresponding eigenvalue. At long distances $x \gg 1$, only the ground state of the transfer matrix Hamiltonian survives. Hence the evolution over infinite distance, defined in Eq. (6), satisfies
\begin{equation}
\label{Hpsi}
 \H \psi_{m}(T) = 0
\end{equation}
and represents the zero mode of the Hamiltonian (\ref{H}).

Let us now fix the normalization for $\psi_m(T)$ using a physical argument. First consider the case $m = 0$, when $\psi_0(T)$ is gauge invariant and can be expressed as a function of $Q = T^{-1} \Lambda T$. If the semi-infinite disordered wire is attached to an ideal metallic lead at the point $x = 0$, the boundary conditions fix $Q(0) = \Lambda$. The overall partition function of the system is $Z = \psi_0(K)$. On the other hand, the supersymmetry requires $Z = 1$, hence $\psi_0(K) = 1$ for any $K$ that commutes with $\Lambda$.  For $m \neq 0$, we can generalize this argument by attaching a perfect metallic lead at $x = 0$ to the infinite wire extended in both $x > 0$ and $x < 0$ directions. Gauge dependence of $\psi_m(T)$ is fixed by the transformation rule (6), that is $\psi_m(KT) = (\sdet K_R)^m f(Q)$. Here the function $f(Q)$ is gauge invariant and independent of the sign of $m$. The supersymmetry condition requires $Z = f^2(\Lambda) = 1$ hence
\begin{equation}
\label{BC}
\psi_m(T = 1) = 1.
\end{equation}

The function $\psi_m(T)$ defined by Eqs.\ (8)--(9) does satisfy the condition (\ref{BC}). Substituting $T = 1$ into Eq.\ (8), we see that $\psi_m(1)$ is the partition function of the supersymmetric sigma model defined on the manifold $\G(n)$. Such sigma models were studied in the context of random matrices in Ref.\ \cite{SIvanov02}. The parameter $m$ corresponds to the number of zero modes in the random matrix while $\kappa$ is related to the energy. The supersymmetry condition of the underlying sigma model requires $\psi_m(1) = 1$ hence Eq.\ (\ref{BC}) is automatically satisfied.

Equation (\ref{Hpsi}) together with the boundary condition (\ref{BC}) has a unique normalizable solution. In the following two sections, we will show that the function given by Eqs.\ (8) and (9) satisfies Eq.\ (\ref{Hpsi}) and thus indeed provides the evolution operator (5).

\subsection{Fierz identities}
\label{Fierz}

Our starting point will be the Lie superalgebra of $\U(n,n|2n)$. An arbitrary element $A$ in this superalgebra can be written as a linear combination of $16n^2$ generators of the algebra $t_i$ as $A = a^i t_i$ (summation over $i$ is implied). The index $i$ runs from $1$ to $16n^2$ with $a^{1,\dots,8n^2}$ commuting variables and $a^{8n^2+1,\dots,16n^2}$ anticommuting variables. The generators $t_i$ will be represented as regular matrices (rather than supermatrices) satisfying
\begin{equation}
k t_i k = 
\begin{cases}
t_i, & 1 \leq i \leq 8n^2, \\
-t_i, & 8n^2+1 \leq i \leq 16n^2,
\end{cases}
\end{equation}
where $k$ is the BF-structure matrix $k = \diag\{1,-1\}_\mathrm{BF}$. 

Now consider a matrix representation of this superalgebra with the generators given by the matrices $(t_i)_{\mu \nu}$ and define the following operator (again an implicit summation over $i$ and $j$)
\begin{equation}
\label{AF_M}
M_{\mu \nu,\sigma \lambda} = \eta^{ij} (k t_i)_{\mu \nu} (k t_j)_{\sigma \lambda}.
\end{equation}
Here $\eta^{ij}$ is the metric on the algebra defined in Eq.\ (\ref{Delta}). The action of the operator (\ref{AF_M}) on an arbitrary element $A_{\mu \nu} = a^i (t_i)_{\mu \nu}$ is given by
\begin{equation}
M_{\mu \nu,\sigma \lambda} A_{\lambda \sigma} = a^l \eta^{ij} (k t_i)_{\mu \nu} \str(t_j t_l) = a^l (k t_l)_{\mu \nu}.
\end{equation}
As a result, $M_{\mu \nu,\sigma \lambda}$ is given by
\begin{equation}
M_{\mu \nu,\sigma \lambda} = \delta_{\nu \sigma} k_{\mu \lambda}.
\end{equation}
This can be used to show that
\begin{subequations}
\label{AF_unit}
\begin{align}
&\eta^{ij} \str(t_i A) \str(t_j B) = M_{\mu \nu, \sigma \lambda} A_{\nu \mu} B_{\lambda \sigma} = \str(A B), \\
&\eta^{ij} \str(t_i A t_j B) = M_{\mu \nu, \sigma \lambda} (A k)_{\nu \sigma} B_{\lambda \mu} = \str A \str B.
\end{align}
\end{subequations}
In the intermediate expressions we also assume summation over repeated lower Greek indices.

Using (\ref{AF_unit}), we can derive similar relations for the generators of the tangent space to any symmetric superspace by applying some additional constraints. For the unitary class, the sigma model manifold is $\U(n,n|2n)/\U(n|n) \times \U(n|n)$. This means that the generators of the tangent space are the generators of the algebra of $\U(n,n|2n)$ that are further restricted to anticommute with the matrix $\Lambda$. Using the condition $\Lambda t_i \Lambda = - t_i$ together with (\ref{AF_unit}), we get the following identities for the unitary class
\begin{subequations}
\label{AF_ClassA}
\begin{align}
&\eta^{ij} \str(t_i A) \str(t_j B) = \frac{1}{2} \Bigl[ \str (A B) - \str(\Lambda A \Lambda B) \Bigr], \label{AF_A_1}\\
&\eta^{ij} \str(t_i A t_j B) = \frac{1}{2}  \Bigl[ \str A \str B - \str(\Lambda A) \str(\Lambda B) \Bigr]. \label{AF_A_2}
\end{align}
\end{subequations}

The corresponding identities for orthogonal and symplectic classes can be obtained from (\ref{AF_ClassA}) by imposing an additional condition that the generators are odd under charge conjugation $\bar{t}_i = C^{T} t_i^T C = - t_i$. Here the charge conjugation matrix $C$ obeys $C^T C = 1$ and $C^2 = k$ ($C^2 = -k$) for orthogonal (symplectic) class. The Fierz identities read
\begin{subequations}
\label{AF_ClassAIAII}
\begin{align}
&\eta^{ij} \str(t_i A) \str(t_j B) = \frac{1}{8} \Bigl[ \str (A - \bar A) (B - \bar B) - \str \Lambda (A - \bar A) \Lambda (B - \bar B) \Bigr], \label{AF_AIAII_1}\\
&\eta^{ij} \str(t_i A t_j B) = \frac{1}{4} \Bigl[ \str A \str B - \str(\Lambda A) \str(\Lambda B) - \str C^2 \bar{A}(B - \Lambda B \Lambda) \Bigr]. \label{AF_AIAII_2}
\end{align}
\end{subequations}

\subsection{Action of the Hamiltonian}

Consider the action of the Laplace-Beltrami operator (\ref{Delta}) on the function $\phi(T)$ defined in Eq. (9)
\begin{align}
 \label{3Z_gammaeta}
 \Delta_Q \phi(T)
  &= \eta^{ij} \frac{\partial}{\partial x^i} \frac{\partial}{\partial x^j} \exp\left[
       -\frac{\kappa}{2\gamma} \str P_{\pm} \bigl( e^{X/2} T + T^{-1} e^{-X/2} \bigr)
     \right]  \biggr|_{X = 0} \nonumber \\
  &= \eta^{ij} \left[
       -\frac{\kappa}{2\gamma} \str t_i t_j (T P_{\pm} + P_{\pm} T^{-1})
       +\frac{\kappa^2}{4 \gamma^2}  \str t_i (T P_{\pm} - P_{\pm} T^{-1}) \str t_j (T P_{\pm} - P_{\pm} T^{-1})
     \right] \phi(T).
\end{align}

The Fierz identities derived in Sec.\ \ref{Fierz} can be now applied to simplify the factor in front of $\phi(T)$. For the first term in square brackets we apply the identity (\ref{AF_A_2}) or (\ref{AF_AIAII_2}) with $B = 1$. Due to the property $\str 1 = \str \Lambda = 0$, this term vanishes identically. The second term in the square brackets can be simplifies with the help of identities (\ref{AF_A_1}) or (\ref{AF_AIAII_1}):
\begin{align}
\label{3Z_A2}
\frac{\kappa^2}{4 \gamma^2} \eta^{ij} \str \bigl[t_i (T P_{\pm} - P_{\pm} T^{-1}) \bigr] \str\bigl[ t_j (T P_{\pm} - P_{\pm} T^{-1}) \bigr]
  &= \frac{\kappa^2}{8 \gamma^2} \str \left[ (T P_{\pm} - P_{\pm} T^{-1})^2 -  (\Lambda T P_{\pm} - \Lambda P_{\pm} T^{-1})^2 \right] \nonumber \\
  &= \frac{\kappa^2}{8 \gamma^2} \str \Lambda T \Lambda T^{-1} = \frac{\kappa^2}{8 \gamma^2} \str \Lambda Q.
\end{align}
Here we have used the properties $P_\pm \Lambda = \pm P_\pm$, $\Lambda = \bar \Lambda$, and $T^{-1} = \bar T$.

We have thus established that
\begin{equation}
 \H \phi(T)
  = \left[ -\frac{\gamma}{2} \Delta_Q + \frac{\kappa^2}{16 \gamma} \str(\Lambda Q) \right] \phi(T)
  = 0.
\end{equation}
Crucially, the prefactor generated by the action of the Laplace-Beltrami operator on $\phi(T)$ depends only on $Q = T^{-1} \Lambda T$. This means that the same equation is satisfied by $\phi(K T)$ for any $[K,\Lambda]=0$. We note here that $\phi(K T)$ already provides a zero mode for the Hamiltonian even before averaging over the gauge group $K$. The averaging performed in Eq.\ (8) is only needed to get a function that transforms properly under gauge transformations (6) and satisfies the normalization condition (\ref{BC}).

\section{Construction of the zero mode}

In this section, we construct the zero mode for the minimal (one replica) model in the unitary, orthogonal, and symplectic classes.

\subsection{Unitary class}

The minimal sigma model for the unitary symmetry class is defined on the manifold $\U(1,1|2)/ \U(1|1) \times \U(1,1)$. For the explicit construction of the zero mode, we fix the gauge by the condition $\psi_m(K T K^{-1}) = \psi_m(T)$ as in Ref.\ \cite{SKhalafA}. This gauge choice implies that the zero mode is independent of $K$ and hence is a function of a single non-compact angle $\theta_B > 0$ and a single compact angle $0 < \theta_F < \pi$. The radial transfer matrix Hamiltonian at finite frequency (\ref{H}) has the form
\begin{equation}
 \H
  = -\frac{1}{J} \left[
      \frac{\partial}{\partial \theta_B} J \frac{\partial}{\partial \theta_B}
      + \frac{\partial}{\partial \theta_F} J \frac{\partial}{\partial \theta_F}
    \right]
    - \frac{m^2}{4} \left[
      \frac{1}{\cosh^2(\theta_B/2)} - \frac{1}{\cos^2(\theta_F/2)}
    \right]
    + \frac{\kappa^2}{8} (\cosh\theta_B - \cos\theta_F), 
 \label{3Z_Hunit}
\end{equation}
with the Jacobian
\begin{equation}
 J = \frac{\sinh \theta_B \sin \theta_F}{(\cosh \theta_B - \cos \theta_F)^2}.
\end{equation}

The representation (8)--(9) for the zero mode involves an integral over a single replica 0D sigma model manifold $\G(1)$ of class AIII [a 1-hyperboloid $H^1 = \GL(1, \mathbb{C})/\mathrm{U}(1)$ in the noncompact sector and a circle $S^1 = \U(1)$ in the compact sector] with $2$ real and $2$ Grassmann variables. An explicit parametrization of this manifold was given in Ref.\ \cite{SIvanov02}. The integral (8)--(9) yields
\begin{align}
\psi_m(\theta_B, \theta_F)
  &= \frac{\kappa}{2} \cosh(\theta_B/2) I_m[\kappa \cos(\theta_F/2)] \Bigl(
      K_{m - 1} [\kappa \cosh(\theta_B/2)] + K_{m + 1} [\kappa \cosh(\theta_B/2)]
    \Bigr) \notag \\
    &\quad + \frac{\kappa}{2} \cos(\theta_F/2) \Bigl(
      I_{m - 1}[\kappa \cos(\theta_F/2)] + I_{m + 1}[\kappa \cos(\theta_F/2)]
    \Bigr) K_m [\kappa \cosh(\theta_B/2)].
 \label{3Z_Psiunit}
\end{align}
Here $I_m(z)$ is the modified Bessel function and $K_m(z)$ is the McDonald function. It is easy to check that this function obeys the equation $\H \psi_m = 0$, with the Hamiltonian (\ref{3Z_Hunit}) and the boundary condition $\psi_m(\theta_F = \theta_B = 0) = 1$ for all values of $m$ and $\kappa$ \cite{SCoulomb}. We can also check that in the limit $\kappa \to 0$, the zero mode of the Laplace-Beltrami operator $[\cos(\theta_F/2)/\cosh(\theta_B/2)]^m$ is recovered, cf.\ Ref.\ \cite{SKhalafA}.

\subsection{Orthogonal class}

The minimal sigma model of the orthogonal class is defined on the manifold $\OSp(2,2|4)/ \OSp(2|2) \times \OSp(2|2)$. The radial transfer matrix Hamiltonian acts on functions with two non-compact angles $\theta_{1,2} > 0$ and one compact angle $0 < \theta_F < \pi$ and is given by
\begin{equation}
 \H = -\frac{1}{J} \left[
        \frac{\partial}{\partial \theta_1} J \frac{\partial}{\partial \theta_1}
        + \frac{\partial}{\partial \theta_2} J \frac{\partial}{\partial \theta_2}
        + \frac{\partial}{\partial \theta_F} J \frac{\partial}{\partial \theta_F}
      \right]
      + \frac{\kappa^2}{8} (\cosh\theta_1 \cosh\theta_2 - \cos\theta_F),
 \label{3Z_Horth}
\end{equation}
with the Jacobian
\begin{equation}
\label{3I_AIJ}
 J
  = \frac{\sinh\theta_1 \sinh\theta_2 \sin^3\theta_F}{[\cosh(\theta_1 + \theta_2) - \cos\theta_F]^2[\cosh(\theta_1 - \theta_2) - \cos\theta_F]^2}.
\end{equation}

The integral representation (8)--(9) for the zero mode involves an integral over a single replica 0D sigma model of class CI [a 1-hyperboloid $H^1 = \SO(2, \mathbb{C})/\SO(2)$ in the noncompact sector and a 3-sphere $S^3 = \Sp(2)$ in the compact sector] with $4$ real and $4$ Grassmann variables. A convenient parametrization for this manifold can be adopted from Ref.\ \cite{SIvanov02}. (In Ref.\ \cite{SIvanov02} a parametrization for the sigma model of class DIII is discussed. It can be adjusted for class CI by interchanging compact and non-compact sectors.) An explicit form of the zero mode is
\begin{align}
\psi_0
  &= \kappa \cosh(\theta_1/2) \cosh(\theta_2/2)\, I_0[\kappa \cos(\theta_F/2)] K_1[\kappa \cosh(\theta_1/2) \cosh(\theta_2/2)] \notag \\
    &\quad + \frac{\kappa (1 + \cosh\theta_1 + \cosh\theta_2 + \cos\theta_F)}{4 \cos(\theta_F/2)}\, I_1[\kappa \cos(\theta_F/2)] K_0[\kappa \cosh(\theta_1/2) \cosh(\theta_2/2)].
 \label{3Z_Psiorth}
\end{align}
It is easy to check that this function indeed obeys the equation $\H \psi_0 = 0$ with the Hamiltonian (\ref{3Z_Horth}) and the boundary condition $\psi_0(\theta_1 = \theta_2 = \theta_F = 0) = 1$.

\subsection{Symplectic class}

The minimal sigma model for the symplectic class is defined on the manifold $\SpO(2,2|4)/\SpO(2|2) \times \SpO(2|2)$. The compact part of this manifold has the structure of the product of two spheres $S^2 \times S^2 / \mathbb{Z}_2$. Factorization over $\mathbb{Z}_2$ implies that simultaneous inversion of both spheres leaves the matrix $Q$ invariant. The manifold has a nontrivial fundamental group $\pi_1 = \mathbb{Z}_2$ hence all trajectories connecting a given pair of points can be classified into two topologically distinct classes. The radial Laplace-Beltrami operator acts on functions with one non-compact angle $\theta_B > 0$ and two compact angles $\theta_1,\theta_2$. If the two compact angles are allowed to vary between $0$ and $\pi$, this will parametrize a full product of two sphere in the compact sector, which is the universal double cover of the sigma model manifold. Functions defined on the sigma model manifold then obey the additional restriction that they are invariant under the simultaneous inversion $\theta_{1,2} \mapsto \pi - \theta_{1,2}$. Functions belonging to the non-trivial topological sector, on the other hand, flip sign under such inversion which is a manifestation of their gauge dependence \cite{SBrouwer96}. 

The transfer matrix Hamiltonian has the form
\begin{equation}
 \H
  = -\frac{1}{J} \left[
        \frac{\partial}{\partial \theta_B} J \frac{\partial}{\partial \theta_B}
        + \frac{\partial}{\partial \theta_1} J \frac{\partial}{\partial \theta_1}
        + \frac{\partial}{\partial \theta_2} J \frac{\partial}{\partial \theta_2}
      \right]
    + \frac{\kappa^2}{8} (\cosh\theta_B - \cos\theta_1 \cos\theta_2),
 \label{3Z_Hsymp}
\end{equation}
with the Jacobian
\begin{equation}
J = \frac{\sinh^3\theta_B \sin\theta_1 \sin\theta_2}{[\cosh\theta_B - \cos(\theta_1 + \theta_2)]^2[\cosh\theta_B - \cos(\theta_1 - \theta_2)]^2}.
\end{equation}

There are two distinct zero-mode functions in the symplectic class which are even/odd under inversion and correspond to even/odd number of channels. The functions are given by the integral (8)--(9) over a single replica 0D sigma model of class DIII [a 3-hyperboloid $H^3 = \mathrm{Sp}(2, \mathbb{C})/\mathrm{Sp}(2)$ in the noncompact sector and $\O(2)$ in the compact sector] with $4$ real and $4$ Grassmann variables. This manifold has two disconnected components, corresponding to the two components of the group $\O(2)$ in the compact sector.  The zero mode in the even/odd sector corresponds to the sum/difference of the integrals over the two disconnected components. We first compute the integral (8) over one of the components taking $\mathrm{SO}(2)$ in the compact sector and using the parametrization of Ref.\ \cite{SIvanov02}. This yields the zero mode function averaged with respect to even/odd number of channels:
\begin{align}
 \psi
  &= \kappa \cos(\theta_1/2) \cos(\theta_2/2)\, I_1[\kappa \cos(\theta_1/2) \cos(\theta_2/2)] K_0[\kappa \cosh(\theta_B/2)] \notag \\
    &\quad + \frac{\kappa (1 + \cos\theta_1 + \cos\theta_2 + \cosh\theta_B)}{4 \cosh(\theta_B/2)}\, I_0[\kappa \cos(\theta_1/2) \cos(\theta_2/2)] K_1[\kappa \cosh(\theta_B/2)].
 \label{3Z_Psi0symp}
\end{align}
The function $\psi$ is not defined on the sigma model manifold, but rather on its double cover, since it is not invariant under simultaneous inversion $\theta_{1,2} \mapsto \pi - \theta_{1,2}$. We construct the zero mode corresponding to the even/odd sectors simply by taking the linear combinations
\begin{equation}
 \psi_{\rm e/o}(\theta_B, \theta_1, \theta_2)
  = \psi(\theta_B, \theta_1, \theta_2) \pm \psi(\theta_B, \pi - \theta_1, \pi - \theta_2).
\end{equation}
Indeed, both functions obey the equation $\H \psi_{\rm e/o} = 0$. The even function is invariant under inversion in the compact sector and thus depends only on $Q$ (as it should be for $m = 0$). The odd function changes sign under inversion which is just a manifestation of its gauge dependence. The squared function $\psi_{\rm e/o}^2$ is gauge invariant in both cases.

\section{Calculation of the return probability}

In this section, we present the details of the calculation of the return probability using the general integral representation of local correlation functions (10). Let us first define the following quantity
\begin{equation}
 B(\kappa)
  = -\frac{1}{16 \gamma^2} \int\limits_{\G(2)} dT\, (\sdet T)^m \str \bigl( k P_+ T^{-1} \Lambda T k P_- T^{-1} \Lambda T \bigr)
    \exp\left[-\frac{\kappa}{2 \gamma} \str \bigl( T + T^{-1} \bigr) \right].
 \label{B}
\end{equation}
The return probability $F(\tau)$ defined in (4) [in the dimensionless units (11)] is then given by the Laplace transform
\begin{equation}
 F(\tau)
  = \lim_{\kappa \to 0} \kappa^2 B(\kappa) - \int_0^\infty \frac{dz}{\pi}\, e^{-\tau z/2} \mathop{\mathrm{Im}} B(\kappa = i \sqrt{z}),
   \label{Ftau}
\end{equation}
which is obtained from (4) by closing the integration contour in the lower half plane of $\omega$. The function $B(\kappa)$ has a branch cut along the negative imaginary axis hence we obtain an integral along this line. In addition, when the topological term is absent, there is a pole at $\omega = -i0$ and the residue determines the saturation value $F(\tau \to +\infty)$.

\subsection{Parametrization of the matrix $T$}

Integration over $T \in \G(2)$ in Eq.\ (\ref{B}) can be performed efficiently using the following parametrization. First, we factor the matrix
\begin{equation}
 \label{3Z_Trg}
 T = T_r T_g,
 \qquad
 T_r = \begin{pmatrix} T_B & 0 \\ 0 & T_F \end{pmatrix}_\text{BF},
\end{equation}
with $T_r$ containing all the commuting variables and $T_g$ parametrizing the Grassmann sector. Due to the group structure of the manifold $\G(2)$, the Jacobian of the above parametrization is unity: $dT = dT_B\, dT_F\, dT_g$. Second, the Grassmann part of the matrix can be parametrized in terms of a general odd element of the algebra $W$ satisfying $k W k = -W$ and $\bar W = - W$ (for orthogonal and symplectic classes) as
\begin{equation}
 \label{3Z_Tg}
 T_g = \sqrt{\frac{1 + W}{1 - W}}.
\end{equation}
This choice leads to a unit Jacobian in the Grassmann sector, $dT_g = dW$, cf.\ Ref.\ \cite{SIvanov06}. Integration over $W$ can be readily performed by expanding the exponential and the pre-exponent in Eq.\ (\ref{B}) in powers of $W$. The remaining integration over $T_r$ is facilitated by choosing the parametrization
\begin{equation}
 \label{3Z_Tr}
 T_r = \begin{pmatrix} V_B^{-1} e^{\hat\theta_B} V_B & 0\\ 0 & V_F^{-1} e^{i \hat\theta_F} V_F \end{pmatrix}_\text{BF}.
\end{equation}
Diagonal matrices $\hat\theta_{B/F}$ parametrize eigenvalues of $T_r$ and contain non-compact ($-\infty < \hat\theta_B < \infty$) and compact ($0 < \hat\theta_F < 2\pi$) angles. The Jacobian of such a parametrization, $d T_r = J(\hat\theta_{B,F})\, d\hat\theta_B\, d\hat\theta_F\, d V_B\, dV_F$, is a trigonometric polynomial in the angles $\hat\theta$.

After Grassmann variables are integrated out, the exponential factor in Eq.\ (\ref{B}) depends only on the eigenvalues $\hat\theta_{B/F}$ while the matrices $V_{B/F}$ enter only the pre-exponent. To facilitate integration over $V_{B/F}$, we will use the Cartan decomposition of these matrices with respect to the matrix $\Lambda = \diag\{1,\, -1\}_\text{RA}$. For both $V_B$ and $V_F$, we write
\begin{equation}
 V = U' A U.
 \label{CartanV}
\end{equation}
Here, the matrices $U'$ and $U$ commute with $\Lambda$ and hence are diagonal in the RA space and $A$ is an abelian matrix whose generators anticommute with $\Lambda$. One advantage of this decomposition is that the integration measure factorizes: $dV = J(A)\, dA\, dU\, dU'$. In addition, since the pre-exponent in Eq.\ (\ref{B}) depends only on the combination $T^{-1}\Lambda T$, some $U$ factors will cancel out.

After the matrices $V$ are integrated out, we are left with an integral over the eigenvalues of $T_r$ that involves a finite trigonometric polynomial of angles $\hat\theta$ and an exponential factor with a simple sum of cosines of individual angles. Hence, the result of the integration is a finite polynomial in modified Bessel functions.

\subsection{Unitary class}

In the unitary class, Eq.\ (\ref{B}) involves an integral over the matrix $T$ that belongs to the two-replica 0D sigma model of class AIII. The matrix $T$ of size $4 \times 4$ operates in the replica (that is equivalent to RA) and Bose-Fermi space and can be parametrized by $8$ real and $8$ Grassmann variables. 

We use the parametrization described above with $T_B \in \GL(2, \mathbb{C})/\U(2)$ and $T_F \in \U(2)$, and write explicitly
\begin{gather}
 T_B
  = V_B^{-1} \begin{pmatrix} e^{\theta_{B1}} & 0 \\ 0 & e^{\theta_{B2}} \end{pmatrix} V_B,
 \qquad
 T_F
  = V_F^{-1} \begin{pmatrix} e^{i \theta_{F1}} & 0 \\ 0 & e^{i \theta_{F2}} \end{pmatrix} V_F, \label{TBF_A} \\
 V_{B,F} = A_{B,F} U_{B,F}, \qquad
 A_{B,F} = \exp\begin{pmatrix} 0 & i\alpha_{B,F}/2 \\ i\alpha_{B,F}/2 & 0 \end{pmatrix}, \qquad
 U_{B,F} = \begin{pmatrix} e^{i\phi_{B,F}} & 0 \\ 0 & e^{-i\phi_{B,F}} \end{pmatrix}.
\end{gather}
The Cartan decomposition (\ref{CartanV}) of the matrices $V_{B,F}$ contains only the factors $A$ and $U$ in this case while $U' = 1$. The integration measure in these variables is given by
\begin{align}
 dT_B &= \sinh^2 \left(\frac{\theta_{B1} - \theta_{B2}}{2} \right)\, d\theta_{B1}\, d\theta_{B2}\, dV_B, & dV_B &=  \sin\alpha_B\, d\alpha_B\, d\phi_B, \\
 dT_F &= \sin^2 \left(\frac{\theta_{F1} - \theta_{F2}}{2} \right)\, d\theta_{F1}\, d\theta_{F2}\, dV_F, & dV_F &= \sin\alpha_F\, d\alpha_F\, d\phi_F \label{dT_A}
\end{align}
up to a constant factor. We specify neither this factor nor the ranges of the variables parametrizing $T$. Instead we will perform integration over all real values of non-compact angles ($\theta_{B1}$, $\theta_{B2}$) and over the interval $[0, 2\pi]$ for compact angles ($\theta_{F1}$, $\theta_{F2}$, $\alpha_{B,F}$, $\phi_{B.F}$) and normalize the final result by the supersymmetry condition $\int dT \exp{[-\str(T+T^{-1})]} = 1$. The same normalization trick we will also apply to the orthogonal and symplectic class below.

With the definitions (\ref{TBF_A})--(\ref{dT_A}), the integration in Eq.\ (\ref{B}) is straightforward and yields
\begin{equation}
 \label{3Z_Bunit}
 B^\mathrm{U}_m(\kappa)
  = \frac{1}{3} \Bigl[ I_{m - 1}(\kappa) K_{m - 1}(\kappa) + 4 I_m(\kappa) K_m(\kappa) + I_{m + 1}(\kappa) K_{m + 1}(\kappa) \Bigr]
    +\frac{4}{3 \kappa} \Bigl[ I_m(\kappa) K_{m - 1}(\kappa) - I_{m + 1}(\kappa) K_m(\kappa) \Bigr].
\end{equation}
After the Laplace transform (\ref{Ftau}), we obtain the return probability given by Eq.\ (12a) in the main text.

\subsection{Orthogonal class}

In the orthogonal class, the calculation involves an integral over the two-replica class CI sigma-model manifold with $16$ real and $16$ Grassmann variables. The matrix $T$ has the size $8 \times 8$ and operates in the replica (that is equivalent to RA), time-reversal, and Bose-Fermi spaces. In addition, it satisfies the charge conjugation constraint $\bar{T} = C^T T^T C = T^{-1}$ with $C^2 = \diag\{1,-1\}_{\rm BF} = k$. We choose the charge conjugation matrix $C$ to be
\begin{equation}
C = \begin{pmatrix} C_B & 0 \\ 0 & C_F \end{pmatrix}_\text{BF},
\qquad
C_B = \begin{pmatrix} 0 & 1 & 0 & 0 \\ 1 & 0 & 0 & 0 \\ 0 & 0 & 0 & 1 \\ 0 & 0 & 1 & 0 \end{pmatrix},
\qquad
C_F = \begin{pmatrix} 0 & 1 & 0 & 0 \\ -1 & 0 & 0 & 0 \\ 0 & 0 & 0 & 1 \\ 0 & 0 & -1 & 0 \end{pmatrix}.
\end{equation}

The matrix $T$ is parametrized according to Eqs.\ (\ref{3Z_Trg})--(\ref{CartanV}). The boson sector $T_B \in \SO(4,\mathbb{C})/\SO(4)$ is a non-compact analog of the group $\SO(4)$. It is a $6$-dimensional manifold of rank $2$ with the angles $\theta_{B1}$, $\theta_{B2}$ parametrizing the eigenvalues and $4$ additional angles for the eigenvectors. Explicitly,
\begin{gather}
 T_B = V_B^{-1} e^{\hat\theta_B} V_B, \qquad
 \hat\theta_B = \diag \Bigl\{ \theta_{B1},\; -\theta_{B1},\; \theta_{B2},\; -\theta_{B2} \Bigr\}, \\
 V_B = A_B U_B, \qquad
 A_B = \exp\left[ \frac{1}{2} \begin{pmatrix}
       0 & 0 & \alpha_1 & \alpha_2 \\
       0 & 0 & \alpha_2 & \alpha_1 \\
       -\alpha_1 & -\alpha_2 & 0 & 0 \\
       -\alpha_2 & -\alpha_1 & 0 & 0
     \end{pmatrix} \right], \qquad
 U_B = \diag\Bigl\{ e^{i\phi_1},\, e^{-i\phi_1},\, e^{i\phi_2},\, e^{-i\phi_2} \Bigr\}. \label{3Z_VB}
\end{gather}
Here again, like in the unitary class discussed above, the Cartan decomposition (\ref{CartanV}) of $V_B$ has $U' = 1$. The volume element in this parametrization is
\begin{equation}
 dT_B = \bigl( \cosh\theta_{B1} - \cosh\theta_{B2} \bigr)^2 d\theta_{B1}\, d\theta_{B2}\, dV_B, \qquad
 dV_B = \sin\alpha_1 \sin\alpha_2\, d\alpha_1\, d\alpha_2\, d\phi_1\, d\phi_2
\end{equation}
up to a constant factor.

The fermion sector $T_F \in \Sp(4)$ is a $10$-dimensional group manifold of rank $2$. It can be conveniently parametrized within the Cartan decomposition (\ref{CartanV}) using quaternion notations,
\begin{gather}
 T_F = V_F^{-1} e^{i \hat\theta_F} V_F, \qquad
 \hat\theta_F = \diag \Bigl\{ \theta_{F1},\; -\theta_{F1},\; \theta_{F2},\; -\theta_{F2} \Bigr\}, \qquad
 V_F = U'_F A_F U_F, \\
 U'_F = \exp \begin{pmatrix} i \bm{\chi}_x & 0 \\ 0 & i \bm{\chi}_x \end{pmatrix}, \qquad
 A_F = \exp \begin{pmatrix} 0 & i\bm{\beta}_z \\ i\bm{\beta}_z & 0 \end{pmatrix}, \qquad
 U_F = \exp \begin{pmatrix} i \bm{\beta}_1 & 0 \\ 0 & i \bm{\beta}_2 \end{pmatrix}, \\
 \bm{\chi}_x = \frac{\chi}{4} \begin{pmatrix} 0 & 1 \\ 1 & 0 \end{pmatrix}, \qquad
 \bm{\beta}_z = \frac{\beta}{2} \begin{pmatrix} 1 & 0 \\ 0 & -1 \end{pmatrix}, \qquad
 \bm{\beta}_{1,2} = \beta_{1,2} \begin{pmatrix}
     \cos\chi_{1,2} & \sin\chi_{1,2}\, e^{-i\eta_{1,2}} \\ \sin\chi_{1,2}\, e^{i\eta_{1,2}} & -\cos\chi_{1,2}
   \end{pmatrix}.
\end{gather}
The integration measure in these variables is given by
\begin{gather}
 dT_F = \bigl( \cos\theta_{F1} - \cos\theta_{F2} \bigr)^2 \sin^2\theta_{F1} \sin^2\theta_{F2}\, d\theta_{F1}\, d\theta_{F2}\, dV_F,\\
 dV_F = \sin^3\beta \sin^2\beta_1 \sin^2\beta_2 \sin\chi \sin\chi_1 \sin\chi_2\,
        d\beta\, d\beta_1\, d\beta_2\, d\chi\, d\chi_1\, d\chi_2\, d\eta_1\, d\eta_2.
\end{gather}

As was discussed previously, the above parametrization guarantees that the integral (\ref{B}) has the form of a polynomial in modified Bessel functions. Explicit calculation yields the surprisingly simple result
\begin{equation}
 \label{3Z_Borth}
 B^\mathrm{O}(\kappa)
  = \frac{4}{3 \kappa^2} \bigl[ 1 + \kappa I_1(\kappa) K_2(\kappa) \bigr] + I_2(\kappa) K_0(\kappa) + I_1(\kappa) K_1(\kappa).
\end{equation}
After Laplace transform (\ref{Ftau}), we obtain the return probability (12b) of the main text.

\subsection{Symplectic class}

Similar to the previous sections, we calculate the correlation function (\ref{B}) for the symplectic class. The matrix $T$ contains $16$ real and $16$ Grassmann variables and belongs to the two-replica sigma-model manifold of class DIII. The base of this manifold contains the non-compact sector $T_B \in \Sp(4, \mathbb{C})/\Sp(4)$, which is analogous to the group $\Sp(4)$, and the compact sector $T_F \in \O(4)$. The manifold has two disconnected components with $\sdet T = \pm 1$. Parametrization of the sector with $\sdet T = 1$ [this corresponds to $T_F \in \SO(4)$] is identical to the orthogonal class discussed above up to switching the compact and non-compact sectors. Separation of real and Grassmann variables and subsequent integration proceeds identically to the orthogonal class leading to
\begin{align}
\label{3Z_Bp}
 B^\text{Sp}_+(\kappa)
  = B^\text{O}(\kappa) - \frac{2}{\kappa^2}  
  = -\frac{2}{3 \kappa^2} \bigl[ 1 - 2 \kappa I_1(\kappa) K_2(\kappa) \bigr] + I_2(\kappa) K_0(\kappa) + I_1(\kappa) K_1(\kappa).
\end{align}

Parametrization for the negative sector $\sdet T = -1$ is slightly different. We will not completely diagonalize $T_F$ but rather decompose it as
\begin{equation}
 T_F = V_F^{-1} \begin{pmatrix} e^{i \theta_F} & 0 & 0 & 0 \\ 0 & e^{-i \theta_F} & 0 & 0 \\ 0 & 0 & 0 & e^{i \alpha} \\ 0 & 0 & e^{-i \alpha} & 0 \end{pmatrix} V_F
\end{equation}
and use Eq.\ (\ref{3Z_VB}) for the matrix $V_F$. This parametrization guarantees $\det T_F = -1$ and has the following measure:
\begin{equation}
 dT_F = \sin^2 \theta_F \sin\alpha_1 \sin\alpha_2\, d\theta_F\, d\alpha\, d\alpha_1\, d\alpha_2\, d\phi_1\, d\phi_2.
\end{equation}

The integral (\ref{B}) over the sector $\sdet T = -1$ has the very simple form
\begin{equation}
\label{3Z_ABm}
B^\text{Sp}_-(\kappa) = \frac{1}{3} K_2(\kappa).
\end{equation}
The result for the symplectic wire with even/odd number of channels is then given by combining (\ref{3Z_Bp}) with (\ref{3Z_ABm}):
\begin{equation}
\label{3Z_Bsymp}
 B^\text{Sp}_\text{e/o}(\kappa)
  = B^\text{Sp}_+(\kappa) \pm B^\text{Sp}_-(\kappa)
  = -\frac{2}{3 \kappa^2} \bigl[ 1 - 2 \kappa I_1(\kappa) K_2(\kappa) \bigr] + I_2(\kappa) K_0(\kappa) + I_1(\kappa) K_1(\kappa) \pm \frac{1}{3} K_2(\kappa).
\end{equation}
The Laplace transform (\ref{Ftau}) yields the return probability (12c) of the main text.


\begin{thebibliography}{99}
\bibitem{footnote1}
In the minimal model (one replica), which is sufficient to compute the average of two Green's functions [Eq.\ (\ref{Wt1})], the matrices $T$ and $Q$ have size $8\times 8$ ($4 \times 4$) in the presence (absence) of TRS.
\bibitem{footnote2}
Strictly speaking, $T$ should belong to a proper Lie supergroup $\tilde{\G}(2n)$ related to $\G(2n)$ in Table \ref{table} by analytic continuation. The group $\tilde{\G}(2n)$ is given by $\U(n,n|2n)$, $\OSp(2n,2n|4n)$ and $\SpO(2n,2n|4n)$ for the unitary, orthogonal, and symplectic classes, respectively.
\end{thebibliography}
\end{document}